\documentclass[
					aps,
               twocolumn,
               prl,showpacs
              ]{revtex4}

\usepackage{amsmath, bbold,bm}
\usepackage{amsfonts}
\usepackage{amssymb}
\usepackage[tight]{subfigure}
\usepackage[dvips]{graphicx}
\begin{document}

\title{Attosecond pulse trains as multi-color coherent control}
\author{J. V. Hern\'andez}
\author{B. D. Esry}
\affiliation{J.R. Macdonald Laboratory, Kansas State University, Manhattan, Kansas 66506}

\begin{abstract}
We present a general description of the interaction between multi-color laser
pulses and atoms and molecules, focusing on the experimentally relevant example of 
infrared (IR) pulses overlapped with attosecond pulse trains (APTs).  This formulation reveals
explicitly and analytically the role of the delay between the IR pulse and APT
as a coherent control parameter.  Our formulation also shows the nearly equivalent
roles of the delay and the carrier-envelope phase in controlling the interference
between different multiphoton pathways.  We illustrate these points by investigating
the single ionization of He and introduce dressed adiabatic hyperspherical potentials to aid 
the discussion.  We confirm the predictions with a full-dimensional, two-electron
solution of the time-dependent Schr\"odinger equation.
\end{abstract}
\date{\today}
\maketitle

At the heart of atomic and molecular physics is the motion of bound electrons.   
Since the relevant time scale for their motion is
on the order of atomic units (1~a.u.$\approx$24~as), detailed studies of 
electron dynamics require forces that act on these time scales.  With such 
forces, an avenue for control over the electron dynamics is also created.  These 
abilities to study and control are the central goals of the nascent field of 
laser-based attosecond physics~\cite{PCork:OPN2008,FKrau:RMP2009}.

Attosecond pulses are generated experimentally via high-harmonic generation, 
either as a single isolated pulse~\cite{MHent:NAT2001} or as an attosecond pulse train 
(APT)~\cite{PPaul:SCI2001}.  The latter, which we will focus on in this Letter, contains 
a large range of harmonics of the fundamental infrared (IR) laser (usually a 
Ti:Sapphire at approximately 800~nm).  With selective filtering, a specific range 
of harmonics --- typically in the extreme ultraviolet range --- can be isolated and 
used to illuminate the target.  Temporally, these harmonics combine to give a 
periodic series of pulses, each of which is an attosecond pulse, thus the label APT.
When overlapped with an IR pulse, an APT permits control through the ability to vary 
the delay between the APT and IR field which can be done on a tens-of-attoseconds scale.  
Experimental control of the ionization probability using such a method has been 
realized~\cite{PJohn:PRL2007,LCock:NJP2009}, as has control over the angular distribution~\cite{OGuye:JPB2005,OGuye:JPB2008}.


Electron dynamics in the combined IR+APT field can, of course, be described by solving 
the time-dependent Schr\"odinger equation~\cite{PJohn:PRL2005}.  To try to gain some physical 
insight, though, both a time-dependent, electric field picture~\cite{PJohn:PRL2007,JMaur:PRL2008} 
and a time-independent, photon based picture~\cite{AMaqu:JMO2007,MMeye:PRA2006} have been 
applied. 
An approach akin to the ``simple man's approximation'' has also been invoked to 
describe the dependence on delay using wave packet replicas~\cite{PRivi:NJP2009}.    

While these attempts to understand the dynamics in IR+APT laser fields have some interpretive 
power, they are not general, exact, nor do they provide a simple route to predicting the outcome of a 
particular experiment.  The numerical solution of the time-dependent Schr\"odinger equation is, in principle, general and 
exact, but is limited in practice to simple systems and cannot be said to provide a simple 
guide to predictions.  Thus, a rigorous description of the problem is desired that can also 
provide a conceptual framework within which predictions can be made simply.

In this Letter, we present just such a theoretical framework.  
This 
formulation is closely related to one we have developed~\cite{VRoud:PRL2007,JHua:JPB2009} to understand the effects 
of the carrier-envelope phase (CEP) for few-cycle pulses and, in fact, demonstrates that both 
the CEP and the IR+APT delay provide essentially identical control over a system.  This 
conclusion follows because both of these parameters can be thought of as controlling 
the relative phase between different multiphoton processes that interfere to produce a given 
physical observable.
In fact, attosecond pulses can be characterized by taking advantage of such interferences~\cite{PPaul:SCI2001,JDahl:PRA2009,NDudo:NP2006}.
These systems are realizations of the canonical coherent control 
scenario proposed by Brumer and Shapiro~\cite{PBrum:ARP1992}.  Moreover, because an APT is a synthesis 
of harmonics, our analysis applies equally well to, and provides a connection to, the 
well-studied two-color experiments on controlling molecules~\cite{PBrum:ARP1992,EChar:PRL1993,EChar:PRL1995,TNguy:PRA2005}.
 

Rather than presenting the most general formulation of our theory --- which, while conceptually 
simple, is notationally cumbersome --- we illustrate it by applying it to the single ionization 
of helium by combined IR+APT fields~\cite{PJohn:PRL2007}.  We confirm our analysis with solutions of the 
full six-dimensional, two-electron, time-dependent Schr\"odinger equation.  In the process, we thus account
for the  modulation in ion yield with delay between 
IR and APT recently observed in Ref.~\cite{PJohn:PRL2007} both analytically and numerically.

Our analytic treatment begins with the time-dependent Schr\"odinger equation in the dipole approximation using the length 
gauge (atomic units are used throughout),
\begin{equation} 
i\frac{\partial}{\partial t}\Psi(t) = 
                \left[H_0 -\vec{\mathcal{E}}(t)\cdot\vec{d}\,\right]\Psi(t), 
\label{TDSE} 
\end{equation} 
where $H_0$ is the field-free Hamiltonian and $\vec{d}$ is the dipole operator.  We  take into 
account the fundamental IR field with frequency $\omega$ and two harmonics of order $n_1$ and $n_2$:
\begin{align} 
\mathcal{E}(t) &= \mathcal{E}_1(t)\cos{(\omega t + \varphi_1)} \nonumber \\ 
&+ \mathcal{E}_{n_1}(t)\cos{\left(n_1\left[\omega (t-\tau)+\varphi_1\right]+\varphi_{n_1}\right)}\nonumber \\ 
&+ \mathcal{E}_{n_2}(t)\cos{\left(n_2\left[\omega (t-\tau)+\varphi_1\right]+\varphi_{n_2}\right)} .
\label{EField} 
\end{align} 
The latter two terms make up the APT in our example.  More harmonics can be added to shorten the 
individual pulses in the APT without any conceptual difficulty, but three colors is the minimum 
needed to describe the most significant effects observed in an IR+APT experiment such as 
Ref.~\cite{PJohn:PRL2007} and simplifies the discussion.    
Note that for our analysis the 
harmonics' pulse envelopes ${\cal E}_{n_i}(t)$, which represent the envelope for the whole pulse 
train, must be independent of the IR-APT delay $\tau$ but are otherwise arbitrary.
Much like the derivation of CEP effects in \cite{JHua:JPB2009}, we note from Eq.~(\ref{TDSE}) 
that $\Psi$ is periodic in $\tau$ as well as in the IR CEP $\varphi_1$.  We can thus expand 
$\Psi$ in both  using a discrete Fourier transform,
\begin{equation} 
\Psi(\tau,\varphi_1;t)=\sum_{ml} e^{im\omega\tau} e^{-il\varphi_1} \psi_{ml}(t) ,
\label{Psi}
\end{equation} 
where $\psi_{ml}(t)$ does not depend on either $\tau$ or $\varphi_1$. Substituting this $\Psi$ 
into Eq.~(\ref{TDSE}) and equating coefficients of the linearly independent exponential functions, we obtain
\begin{align} 
i \frac{\partial}{\partial t}\phi_{ml} =& (H_0-l\omega) \phi_{ml} 
  -\frac{1}{2}\vec{\cal E}_{1}\!\cdot\vec{d}\left(\phi_{m,l+1}+\phi_{m,l-1}\right) \nonumber \\ 
  -\frac{1}{2}\vec{\cal E}_{n_1}\!\cdot&\vec{d}\left(e^{i\varphi_{n_1}} \phi_{m+n_1,l+n_1}\!+\!e^{-i\varphi_{n_1}}\phi_{m-n_1,l-n_1}\right)\nonumber  \\ 
  -\frac{1}{2}\vec{\cal E}_{n_2}\!\cdot&\vec{d}\left(e^{i\varphi_{n_2}} \phi_{m+n_2,l+n_2}\!+\!e^{-i\varphi_{n_2}}\phi_{m-n_2,l-n_2}\right) .
\label{FTDSE}
\end{align} 
In these equations, we have made the additional substitution $\psi_{ml}(t)=e^{-il\omega t}\phi_{ml}(t)$ 
to emphasize the connection with the Floquet representation: in the limit that the pulse lengths 
go to infinity, Eq.~(\ref{FTDSE}) reduces to many-mode Floquet theory~\cite{SIChu:PR2004}.  We stress that 
Eq.~(\ref{FTDSE}) is exact even for few-cycle pulses.  We could further remove 
the harmonic phases $\varphi_{n_i}$ with Fourier transforms, but that will not be necessary for the 
current discussion.

We have rigorously obtained a representation in which IR+APT experiments can be understood.  
Further, Eq.~(\ref{Psi}) shows that the CEP and IR-APT delay provide similar  
control over the dynamics of a system. The physical interpretation of this control as interference of 
different photon pathways stems from the facts that the components $\phi_{ml}$ oscillate with 
time-dependence $e^{il\omega t}$ and correlate to the photon components in the Floquet limit.  
Since we consider  photons with different frequencies, the index $l$ represents only the net 
energy $l\omega$ of the photons involved, and can be written as
$l$=$l_1\!+\!l_{n_1} n_1\!+\!l_{n_2}n_2$
with $l_i$ the number of photons of each color.  

Specializing the above discussion to helium determines $H_0$ and $\vec{d}$ in 
Eq.~(\ref{FTDSE}).  We  treat the full 
three-dimensional motion of both electrons using the adiabatic 
hyperspherical representation, which has proven especially useful in the past for understanding the 
dynamics of He~\cite{CDLin:AAM1986,XGuan:PRA2006}.  More specifically, we expand each component $\phi_{ml}$ as 
\begin{equation}
\phi_{ml}=\sum_{L\nu} G_{mlL\nu}(R,t)\Phi_{L\nu}(R;\Omega)
\label{HyperExpansion}
\end{equation}
where $R$ is the hyperradius and $\Omega$ the five hyperangles.
The channel functions $\Phi_{L\nu}$ solve the field-free fixed-$R$ adiabatic equation
$H_{\rm ad} \Phi_{L\nu}$=$U_{L\nu}\Phi_{L\nu}$
for a given total orbital angular momentum $L$ and thus form a complete set at each $R$ indexed by 
$\nu$.  We use Delves' hyperangle and treat the system in the body frame, expanding the Euler-angle dependence of $\Phi_{L\nu}$ 
on Wigner $D$-functions (see \cite{CDLin:AAM1986} for details).  The adiabatic equation 
thus reduces to coupled two-dimensional partial differential equations that we solve using b-splines~\cite{CdeBO:BOOK}.

\begin{figure}
	\includegraphics[width=0.8\linewidth]{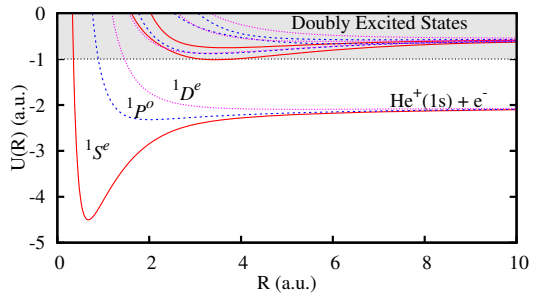}
	\caption{(Color online) Adiabatic hyperspherical potential curves for He.  Shown are $L=0,1,2$ and the region of double excitation.}
	\label{BarePots}
\end{figure}

Upon substituting $\phi_{ml}$ from (\ref{HyperExpansion}) into Eq.~(\ref{FTDSE}) and projecting out the 
$\Phi_{L\nu}$
we find coupled, time-dependent equations for $G_{mlL\nu}(R,t)$ describing laser-driven motion on the adiabatic potentials $U_{L\nu}$. 
Since the experiments in \cite{PJohn:PRL2007} considered only single ionization at energies well below 
doubly-excited resonances, we can to a good approximation  consider only the lowest singlet adiabatic 
potential $\nu$=1 for each $L$.  These potentials support the singly-excited states $1snL$~$^1L$, and 
several examples are shown in Fig.~\ref{BarePots}.  
With this truncation of the $\nu$ expansion, we find a ground state energy of --2.895~a.u. 

One advantage of using the adiabatic hyperspherical representation in the Floquet-like equations 
(\ref{FTDSE}) is that the resulting dressed potentials ---
which include the effects of both Coulomb interactions and the laser field --- 
give insight into the electronic dynamics in much the same way that Floquet potentials 
have for the nuclear dynamics of molecules.  Applied to H$_2^+$, Floquet potentials 
have provided natural explanations of bond-softening, above-threshold dissociation,  vibrational 
trapping, and zero-photon dissociation~\cite{JPost:RPP2004,XHe:PRA1990}.  Previous generalizations of 
the Floquet picture for H$_2^+$ have even helped uncover unexpected phenomena such as above-threshold 
Coulomb explosion~\cite{BEsry:PRL2006}.  It would be interesting to explore the analogues of these 
mechanisms for He in the present representation, but we will focus here on their application to the IR+APT case.

\begin{figure}
	\includegraphics[width=0.8\linewidth]{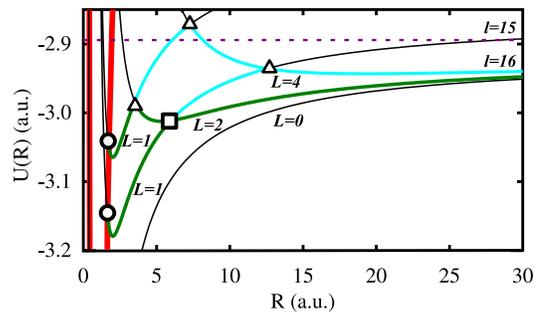}
	\caption{(Color online) Diabatic dressed potentials for He in a three color field ($\omega$, 13$\omega$, and 15$\omega$).  
	Only the most significant curves and crossing are shown, 
	where a circle
	indicates a single large-photon crossing; a square a 1-IR photon crossing; and a triangle, a 3-IR photon 
	crossing. The dashed line indicates the energy of the system, which is approximately conserved
	in this representation.}
	\label{DressedPots}
\end{figure}

Figure~\ref{DressedPots} shows the diabatic dressed potentials assuming
$\omega$ corresponds to 800~nm and the harmonics are $n_1$=13 and $n_2$=15
(the two most 
intense components of the APT in \cite{PJohn:PRL2007} by a factor of at least 5).  These curves are 
generated by first noting that the initial state of the system, $1s^2$~$^1S^e$, has $(m,l)$=(0,0).  
Then, per Eq.~(\ref{FTDSE}), this state couples to the $^1P^o$ channel with one IR photon (0,$\pm$1) 
or with one harmonic photon ($\pm n_i$,$\pm n_i$).  The $^1P^o$ potential in each case is added to the 
figure shifted by $l\omega$.  Each of these new curves can couple to either $^1S^e$ or $^1D^e$ symmetries 
according to dipole selection rules, shifted in energy by an appropriate amount.  Repeating 
this procedure, all possible multiphoton pathways combining IR and harmonic photons to any particular final 
state can be generated as shown in Fig.~\ref{DressedPots}.  
The energy in such a figure can be thought of as being approximately conserved, so that the 
dynamics 
reduces to the well-understood problem of multichannel dynamics.  
This picture, for example, 
maps ionization onto predissociation of 
molecules in the Born-Oppenheimer approximation or, equivalently, onto atomic autoionization in the 
field-free adiabatic hyperspherical representation.
More specifically, transitions are most likely to occur where potentials cross, and  
crossings of curves differing by the fewest photons will dominate.
In fact, diagonalizing the 
right-hand-side of Eq.~(\ref{FTDSE}) at fixed $R$ will produce adiabatic Floquet-like potentials in which 
these crossings become avoided, thus incorporating the strength of the field in the potentials themselves.  
The remaining time dependence 
in ${\cal E}_{i}$ modulates the strength of the 
coupling --- or the gap at the avoided crossings --- as a function of time.    

For the parameters above, single ionization of He requires the absorption of at least 16$\omega$ worth 
of energy, i.e. 16 IR photons, 1 13th harmonic plus 3 IR photons, or 1 15th harmonic plus 1 IR photons.  
Since it requires the fewest photons, we expect the latter scenario to be most likely, leaving the 
system with $(m,l)$=(15,16) and $L$=0 or 2 based on dipole selection rules.  All together, we
would write these pathways as (0,0)$^1S^e$$\rightarrow$(15,15)$^1P^o$$\rightarrow$(15,16)$^1S^e$ or $^1D^e$.
Following the dressed potentials 
in Fig.~\ref{DressedPots} suggests that only $L$=2 will be populated substantially, however, since it crosses 
the (15,15)$^1P^o$ curve and $L$=0 does not.  
The pathway beginning with
the 13th harmonic
would mainly proceed via 
(0,0)$^1S^e$$\rightarrow$(13,13)$^1P^o$$\rightarrow$(13,16)$^1D^e$ by similar arguments.  The above two paths are
highlighted in green in Fig.~\ref{DressedPots}.
Another set of likely pathways (light blue in Fig.~\ref{DressedPots}) ends on 
$^1$G$^e$:$\;(0,0)^1S^e$$\rightarrow$(15,15)$^1P^o$$\rightarrow$(15,16)$^1G^e$ and
(0,0)$^1S^e$$\rightarrow$(13,13)$^1P^o$$\rightarrow$(13,16)$^1G^e$.   
Other outcomes are certainly possible and will occur with some 
probability.  The goal of this kind of analysis, though, is to try to identify the main routes to a 
given final state.
Since there exists more than one pathway to this lowest above-threshold ionization (ATI) peak, interference 
will occur.   Similar considerations show that there are generally multiple paths to ionization
for each higher ATI peak 
$l$=17,18,19,\ldots  
opening up the possibility of 
control.

In this manner, our formulation of the problem lets us learn a considerable amount about 
controlling this interference without solving the time-dependent problem.  
To see explicitly how these pathways interfere, we can derive an expression for any physical observable using Eq.~(\ref{Psi}).  
In particular, for the He IR+APT photoelectron spectrum discussed above, we obtain
\begin{multline}
 \frac{dP_{\rm ion}}{dE} 
  \!= \!\sum_{L}\sum_{m,m'}\sum_{l,l'} e^{i(m-m')\omega\tau} e^{i(l'-l)\varphi_1}  \\
  \times 
\langle E L|F_{mlL}\rangle
\langle E L|F_{m'l'L}\rangle^*,
\label{eqn04}
\end{multline}
where $|EL\rangle$ is an energy-normalized continuum state with asymptotic kinetic energy $E$ and
$F_{mlL}$=$e^{-il\omega t}G_{mlL}$.  All of the $\tau$ and $\varphi_1$ dependence is
displayed analytically since the amplitudes $\langle E L|F_{mlL}\rangle$ are
independent of these parameters, determining only which pathways will interfere.
Equation~(\ref{eqn04}) shows that the delay dependence is determined only by the difference
in $m$, essentially the harmonic order, between two pathways.  For the lowest ATI peak discussed above, the 
final states of two of the pathways identified were (13,16)$^1D^e$ and (15,16)$^1D^e$ for the 13th and 15th
harmonics, respectively.  Since these have the same final $l$ (net energy absorbed
from the field), 
they will contribute at the same final energy
and interfere, resulting in a sinusoidal dependence on $\omega\tau$ with period
$\pi$ since $|m$--$m'|$=2.  
This $\pi$ periodicity in the ionization yield is exactly what was observed
in both the experiment and numerical results presented in \cite{PJohn:PRL2007,LCock:NJP2009}.
In general, the periodicity in $\omega\tau$ will be given by $2\pi/|m\!-\!m'|$.
Thus, if even harmonics had also been included, the periodicity would have been
2$\pi$ since $|m$--$m'|$ would equal 1.
To see the 2$\pi$ periodicity, however, one would have to measure a quantity that -- unlike the
energy spectrum (\ref{eqn04}) -- takes into accout interference between different final $L$s, such as the 
momentum distribution.

\begin{figure}
		\includegraphics[width=0.5\linewidth]{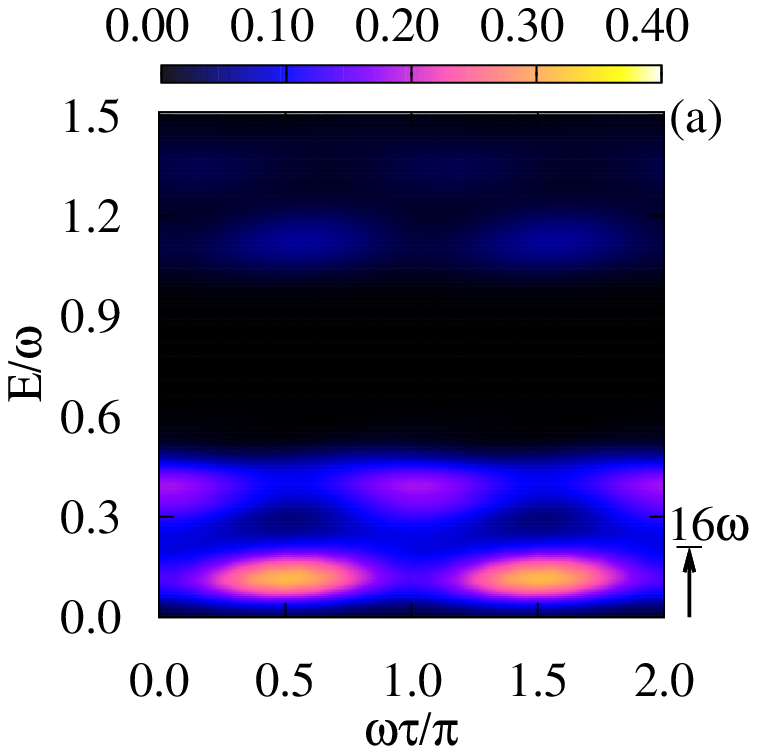}
		\includegraphics[width=0.4\linewidth]{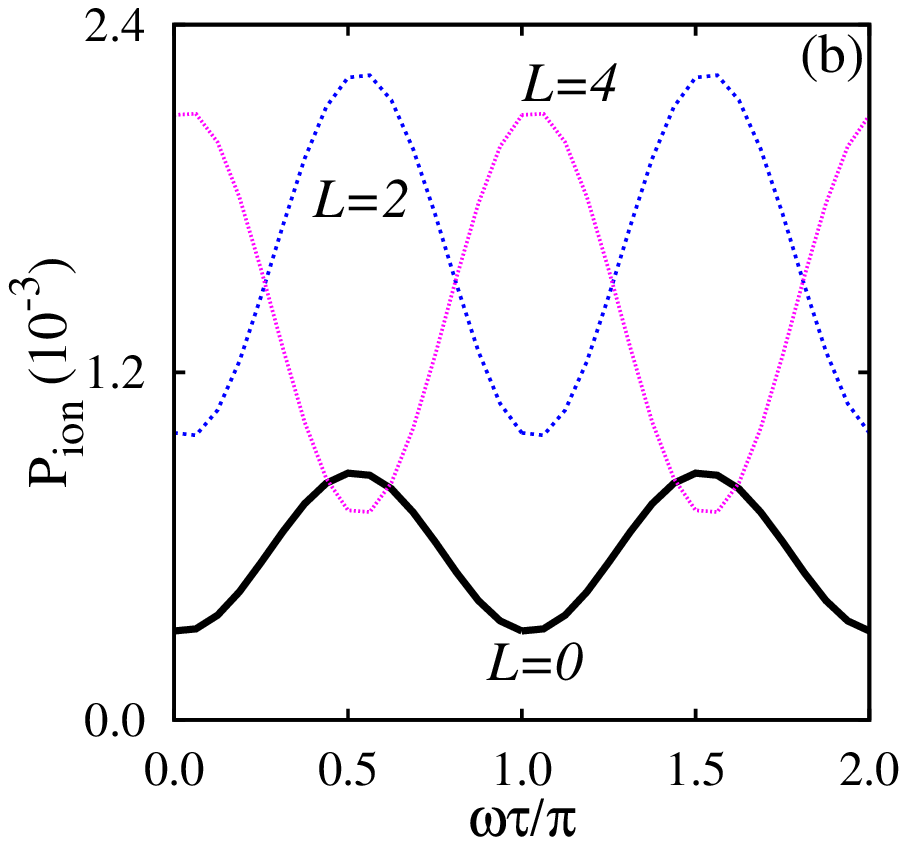}
\caption{(Color online) (a) Calculated photoelectron spectrum 
				for an 800 nm IR pulse with 
				$I_1=5.0\times 10^{12}$ W/cm$^2$ and $I_{13} =I_{15}=0.1I_1$. 
			(b) Ionization probability as a function of delay for $L$=0, 2, and 4.  
}
\label{PESpectrum}
\end{figure}
To confirm our analysis, we carried out a full-dimensional numerical solution of the time-dependent Schr\"odinger equation
 (\ref{TDSE}) using the adiabatic hyperspherical representation.
Figure~\ref{PESpectrum}(a) shows the resulting photoelectron spectrum, again for parameters chosen to be similar to
the experiment in \cite{PJohn:PRL2007}.
As expected, the calculations show ATI peaks separated by $\omega$.  The largest peak is the lowest one (indicated by the arrow), 
which was discussed above,
and the predicted $\pi$ period of its modulation is clearly seen.  
A closer look at the 16$\omega$ peak reveals
that it is split in energy with the lower-energy peak primarily $L$=2 and the higher-energy peak $L$=4 [see Fig.~\ref{PESpectrum}(b)]. 
These details, however, only serve to bolster the value of the dressed potentials 
we introduced in Fig.~\ref{DressedPots} since 
they allowed us to predict that the
16$\omega$ peak should be dominated by $L$=2 and 4.
The time-dependent calculations and the dressed
potential picture complement each other, providing both quantitative and qualitative understanding
of these processes.

This discussion hardly exhausts the insight that can be gleaned from Eq.~(\ref{eqn04}).
Another example is the fact that the delay $\tau$ between IR and APT will not generate
a left-right asymmetry along the laser polarization
since all of the $L$s corresponding to a given ATI peak are either even or odd (with only odd harmonics in the APT).  
An exception
to this statement occurs for broad bandwidth pulses since the interference between even and
odd $L$s necessary for asymmetry can be produced in the overlap of neighboring ATI peaks.  
Of course, the CEP will also contribute to the asymmetry in this case.

In this Letter, we have presented a general, exact representation for the 
interaction of multicolor laser pulses with atoms and molecules upon which 
a relatively simple interpretive, predictive picture can be built.  In the process,
we introduced
dressed potentials for He --- again based on a rigorous derivation --- that
confer many of the benefits such potentials have provided for molecules on the
atomic problem.
We have also shown that control via the CEP and the delay between multicolor fields, IR+APT
in particular, are conceptually
equivalent, being the result of interference between different multiphoton 
pathways between the initial and final states.  Experiments exercising this
control are thus quite properly thought of as coherent control.


\begin{acknowledgments}
This work is supported by the Chemical Sciences, Geoscience, and Biosciences Division, 
Office for Basic Energy Sciences,Office of Science, U.S. Department of Energy.
\end{acknowledgments} 


\end{document}